\documentclass[aps,prl,twocolumn,showpacs,superscriptaddress]{revtex4}
\usepackage{amsmath}
\usepackage{dcolumn}
\usepackage{epsfig}
\usepackage{graphicx}
\usepackage{latexsym}

\def\S2RO4{Sr$_2$RuO$_{4}$}
\def\SRO3{SrRuO$_{3}$}
\def\STO3{SrTiO$_{3}$}

\begin{document}
\hyphenation{Ka-pi-tul-nik}



\title{High Resolution Polar Kerr Effect Measurements of \S2RO4: Evidence for  Broken Time Reversal Symmetry  in the Superconducting State}

\author{Jing Xia}
\affiliation{Department of Physics, Stanford University, Stanford, CA 94305}
\author{ Yoshiteru Maeno}
\affiliation{Department of Physics, Kyoto University, Kyoto 606-8502, Japan}
\author{Peter T. Beyersdorf}
\affiliation{Department of Physics and Astronomy, San Jose State University, San Jose, CA 95192}
\author{ M. M. Fejer } 
\affiliation{Department of Applied Physics, Stanford University, Stanford, CA 94305}
\author{Aharon Kapitulnik}
\affiliation{Department of Physics, Stanford University, Stanford, CA 94305}
\affiliation{Department of Applied Physics, Stanford University, Stanford, CA 94305} 

\date{\today}

\begin{abstract}
Polar Kerr effect in the spin-triplet superconductor \S2RO4 was measured with high precision using a Sagnac interferometer with a zero-area Sagnac loop. We observed non-zero Kerr rotations as big as 65 nanorad appearing below $T_c$ in large domains. Our results imply a broken time reversal symmetry state in the superconducting state of \S2RO4,  similar to $^3$He-A.\end{abstract}


\maketitle

Soon after the discovery of the layered-perovskite superconductor  \S2RO4 \cite{maeno},  it was predicted to be an odd-parity superconductor \cite{rice1,baskaran}. Subsequently, a large body of experimental results in support of odd-parity superconductivity has been obtained \cite{mackenzie}, with the most recent one being a phase-sensitive measurement \cite{nelson}. The symmetry of the superconducting state is related simply to the relative orbital angular momentum of the electrons in each Cooper pair. Odd parity corresponds to odd orbital angular momentum and symmetric spin-triplet pairing. While a priori the angular momentum state can be  $p$ (i.e. $L=1$), $f$ ( i.e. $L=3$), or even higher order \cite{annett,sigrist}, theoretical analyses of superconductivity in \S2RO4 favor the $p$-wave order parameter symmetry \cite{rice1,machida}.  There are many allowed $p$-wave states that satisfy the cylindrical Fermi surface for a tetragonal crystal which is the case of \S2RO4 (see e.g. table IV in \cite{mackenzie}). Some of these states break time-reversal symmetry (TRS), since the condensate has an overall magnetic moment because of either the spin or orbital (or both) parts of the pair wave function.  While an ideal sample will not exhibit a net magnetic moment,  surfaces and defects at which the Meissner screening of the TRS-breaking moment is not perfect can result in a small magnetic signal \cite{sigrist}.  Indeed, muon spin relaxation ($\mu$SR) measurements on good quality single crystals of \S2RO4 showed excess relaxation that spontaneously appear at the superconducting transition temperature.  The exponential nature of the increased relaxation suggested that its source is a broad distribution of internal fields, of strength $\sim$ 0.5 Oe, from a dilute array of sources \cite{luke1,luke2}. While TRS breaking is not the only explanation for the $\mu$SR observations, it was accepted as the most likely one \cite{mackenzie}. However, since the existence of TRS breaking has considerable implications for understanding the superconductivity of \S2RO4, establishing the existence of this effect, and in particular in the bulk without relying on imperfections and defects is of utmost importance. The challenge is therefore to couple to the TRS-breaking part of the order parameter to demonstrate the effect unambiguously.

In this paper we show results of polar Kerr effect (PKE) measurements on high quality single crystals of \S2RO4. In these measurements we are searching for an effect analogous to the magneto-optic Kerr effect (MOKE) which would cause a rotation of the direction of polarization of the reflected linearly polarized light normally incident to the superconducting planes. PKE is sensitive to TRS breaking since it measures the existence of an antisymmetric contribution to the real and imaginary parts of the frequency-dependent dielectric tensor. Such a contribution is necessarily absent if TRS is not broken in the material. Our results show unambiguously the emergence of a finite, PKE at $T_c \approx 1.5$ K.  The size of the effect increases with decreasing temperature down to $0.5$ K, tracking the increase in magnitude of the superconducting order parameter. Combining our result with previously published results pertaining to the properties of the superconducting order parameter in \S2RO4, we conclude that the order parameter is {\bf d(k)}=$\hat{\bf z}[k_x \pm ik_y]$, where using the convention of Balian and Werthamer \cite{balian} we used the three-component complex vector {\bf d(k)} to represent the superconducting gap-matrix. To detect the very small PKE ($\sim$ 65 nanorad) we used a new scheme based on a zero-area-loop polarization-Sagnac interferometer at wavelength of $\lambda =$1550 nm.  The use of Sagnac interferometry to characterize TRS breaking was first introduced by our group for the search for ``anyon superconductivity" \cite{spielman1,spielman2}. The new scheme is $\sim 20$ times more sensitive, reaching shot-noise limit of 100 nanorad/$\sqrt{\rm Hz}$ at 10 $\mu$W of detected optical power from room temperature down to 0.5 K \cite{xia}.

\begin{figure}
\begin{center}
\includegraphics[width=0.95 \columnwidth]{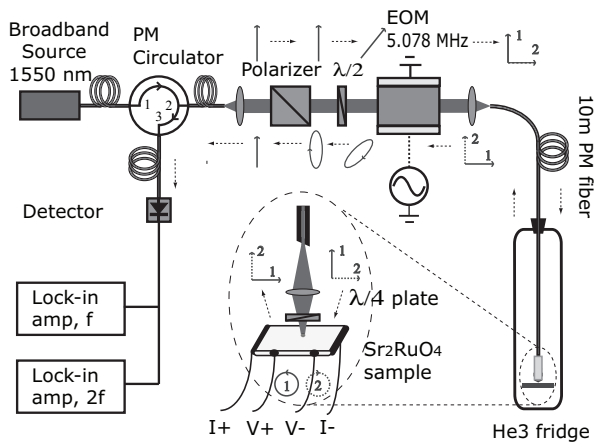}
\end{center}
\vspace{-4mm}
\caption{ Experimental setup and the polarization states at each locations: vertical, in-plane polarization; horizontal, out-of-plane polarization; solid line, beam 1; dashed line, beam 2. } 
\label{setup}
\end{figure}

Before introducing our results, we describe briefly the novel apparatus that we built for the PKE measurements and its ability to probe non-reciprocal circular birefringence effects, while rejecting reciprocal effects to unprecedented accuracy.  Fig.~\ref{setup} shows a schematic of the design. The output of a very-short-coherence length ($\sim$30 $\mu$m) fiber-coupled superluminescent light emitting diode (SLED) centered at 1550 nm is routed by a fiber polarization-maintaining (PM) circulator to a Glan-Calcite polarizer, and becomes linearly polarized. The polarization is then rotated by a half-wave ($\lambda /2$) plate at 45$^o$  to the axes of a birefringent electro-optic modulator (EOM), which operated at $f_m = 5.078$ MHz, corresponding to twice the optical transit time of the system. After passing the EOM, the beam is split into two parts, with polarizations along the fast and slow axis of the EOM. They are then launched into the fast and slow axis respectively of a 10-m-long polarization maintaining (PM) fiber that is fed into a He-3 cryostat (base temperature $<$ 0.5 K). An aspheric lens is used to focus the light coming out of the fiber through a 100-$\mu$m-thick quartz quarter-wave ($\lambda /4$) plate into a spot with $1/e^2$ diameter of $\sim$25 $\mu$m on the surface of the \S2RO4 sample. The $\lambda /4$ plate is aligned at 45$^o$  to the axis of the PM fiber and converts the two orthogonally polarized beams into right- and left-circularly polarized light. The non-reciprocal phase shift $\phi_{nr}$ between the two circularly polarized beams upon reflection from the TRS-breaking sample is double the Kerr rotation \cite{spielman1,spielman2,kdf} ($\phi_{nr} = 2 \theta_K$). The same $\lambda /4$ plate converts the reflected beams back into linear polarizations, but with a net 90$^o$ rotation of polarization axis. In this way, the two beams effectively exchange their paths when they travel back through the PM fiber and the EOM to the polarizer. After passing the polarizer, the light is routed by circulator to an AC-coupled photo-detector. Therefore the two main beams travel precisely the same distance from source to detector, except for a small phase difference $\phi_{nr}$, which is solely from the TRS-breaking sample. On the other hand, the distance traveled by light which did not follow the correct path (i.e. due to back reflections and scattering, as well as polarization coupling due to birefringence of the sample or imperfections and misalignment of waveplates) will differ by many times the coherence length due to the birefringence of the PM fiber and EOM. This light may reach the detector, but can't interfere coherently with the main beams; it will at most add a constant background. The EOM serves as a convenient way to actively bias the interferometer to its maximum response and enable lock-in detection \cite{spielman1,spielman2}. The signal from the detector will contain even harmonics of $f_m$ proportional to the overall reflected intensity, and odd harmonics proportional to $\phi_{nr}$\cite{spielman1,spielman2}. Details of (the previous version of) this apparatus, including its low temperature performance, was described in detail by Xia {\it et al.} \cite{xia}. Unlike the previous version, here we removed the focusing lens between the $\lambda /4$ plate and the sample in order to minimize the drift due to the Faraday effect in that lens. The drift of the system is typically $<$ 50 nanorad over 24 hours.

The \S2RO4 samples used were single crystals, grown by a floating-zone method \cite{mao1}. $T_c$ as determined by bulk ac susceptibility measurements on pieces from the same crystal bar was 1.44 K with a transition width of 30 mK. This $T_c$ agrees with the resistive measurement performed in-situ on the measured sample as explained below.  The samples used for the measurements were approximately 3$\times$3 mm$^2$ in area (ab-plane) and 0.25 mm thick (along $c$-axis). $H_{c1}$ of these samples \cite{deguchi}, taking into account demagnetization effects is  5 - 10 Oe. 

For the TRS breaking study in \S2RO4 extra care has to be taken to prevent heating. To this end the sample was mounted on a copper platform with Au wires connected to its two opposite sides and two other Au wires connected to a third side. Besides providing good thermal anchoring, this arrangement was also used for in-situ resistance measurements to verify the transition temperature. Incident optical power on the sample was between 1.8 and 2 $\mu$W during all the measurements. The optical power absorbed by the sample was less than 0.8 $\mu$W due to the good reflectivity ($\sim$60$\%$) \cite{katsufuji} at 1550 nm wavelength, and the detected optical power was $\sim$ 0.6 $\mu$W. With a focused spot diameter of $\sim$25 $\mu$m, the maximum temperature increase due to optical heating was calculated to be less than 100 mK at 0.5 K and 30 mK at $T_c$ respectively, using material parameters given in \cite{mackenzie}.

\begin{figure}[h]
\begin{center}
\includegraphics[width=0.95 \columnwidth]{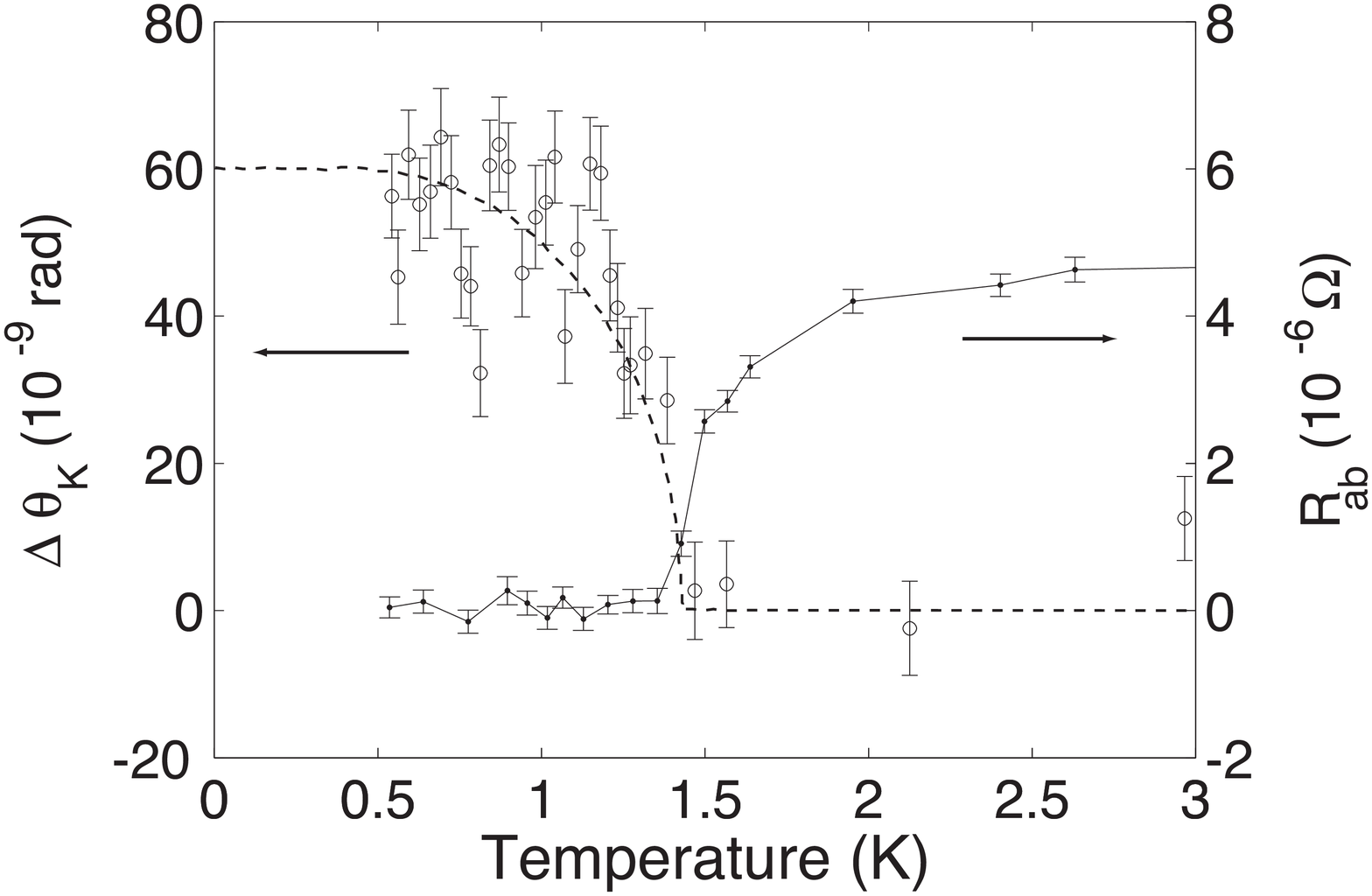}
\end{center}
\vspace{-4mm}
\caption{ Zero-field (earth field) measurement of Kerr effect (circles) and ab-plane electrical resistance (dots). Dashed curve is a fit to a BCS gap temperature dependence.} 
\label{fig2}
\end{figure}

Fig.~\ref{fig2} shows the polar Kerr signal of a sample that was cooled in zero field (more correctly, the field at the location of the sample was measured to be $<$ 0.2 Oe). Data was collected upon warming up and each data point represents a time-average of 1600 seconds. Error bars represent statistical uncertainty.  In all figures showing Kerr effect data we plot $\Delta \theta_K(T) = \theta_K(T) - \theta_K^n$, where $\theta_K^n=\theta_K({\rm 3 K})$ is the normal-state baseline Kerr angle representing the offset of the instrument in zero field (typically $<$ 200 nanorad), or the Kerr angle of the instrument in the presence of a magnetic field. Dashed curve is a fit to a BCS gap temperature dependence.  Also shown in this figure is the resistive transition recorded for the same sample.  In general we did not measure the two quantities simultaneously to avoid possible effects of the current on the measurement.

The data presented in Fig.~\ref{fig2} show a clear increase in $\theta_K$ below $T_c$.  With decreasing temperature, the signal increases sublinearly and seems to saturate to a value of 60 $\pm$ 10 nanorad.  The fact that the size of the signal fluctuates can be due to transient effects in the sample, for example due to occasional vortices which attempt to modify the sense of chirality in the sample (no signal can originate from vortices as we will demonstrate below).  In 6 zero-field runs in different parts of the sample, we measured positive Kerr phase shifts in three runs, negative Kerr phase shifts in two runs, and reduced Kerr phase shift in which the signal changes sign as the sample is warmed up from 0.5 K in one run.  This suggests that if there are domains in the sample they are a few times larger than the beam size (possibly $\sim$ 50 - 100 $\mu$m).

\begin{figure}[h]
\begin{center}
\includegraphics[width=0.95 \columnwidth]{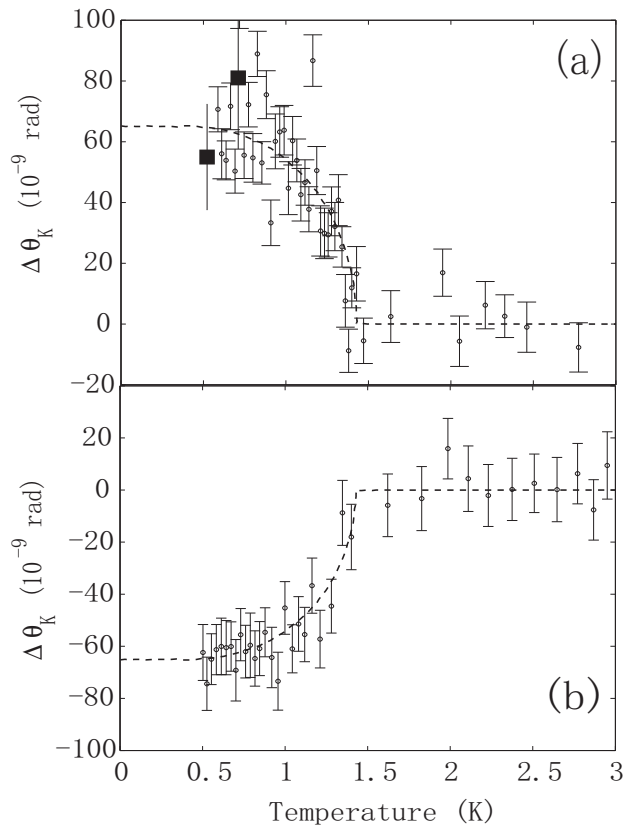}
\end{center}
\vspace{-4mm}
\caption{Representative results of training the chirality with an applied field.  a) + 93 Oe field cool, then zero field warm-up (circles). The two solid squares represent the last two points just before the field was turned off. b) - 47 Oe field cool, then zero field warm-up (circles).  Dashed curves are fits to a BCS gap temperature dependence.}
\label{fig3}
\end{figure}

The broken-TRS is expected to have two possible chiralities.  To choose between the two possible states, a TRS-breaking field such as a magnetic field, that couples to the order parameter can be applied.    Fig.~\ref{fig3}a shows the zero-field warm-up PKE measurement after the sample was cooled through $T_c$ in a field of +93 Oe, while Fig.~\ref{fig3}b shows the zero-field warm-up PKE measurement after the sample was cooled through $T_c$ in a field of -47 Oe. Clearly the two curves give a similar signal below $T_c$ which is equal in magnitude and opposite in sign. This is a clear indication that the applied field indeed influenced the direction of the chirality.  Furthermore, the fact that the size of the effect and its temperature dependence are  the same as in zero-field cooled experiments indicate that the signal we observe is not due to trapped flux. Also, the temperature dependence is clearly seen here to follow a BCS gap function.  Measurements at -93 Oe or at +47 Oe produced similar results to Fig.~\ref{fig3} but with opposite sign. An even stronger indication that no trapped flux is involved is shown in Fig.~\ref{fig3}a where we mark the last two points of the Kerr angle just before the +93 Oe field was turned off to zero.  Clearly no excess signal exists in the presence of the field. Moreover, in a field of +93 Oe there are  $\sim$ 3,000 vortices. If the observed signal was entirely due to vortices, it would imply a single vortex Kerr rotation of less than $10^{-11}$ rad, much below our sensitivity. Thus we proved that a few possible trapped vortices \cite{moler} could not give the observed signal at zero field.  Finally, we observed that while fields that are clearly above $H_{c1}$ could be used to order the sense of chirality,  measurements in which we cooled the sample in fields as high as 4.7 Oe gave random sign of the PKE, similar to zero field (with the same $\sim$ 65 nanorad magnitude).   It is therefore reasonable to assume that ordering fields need to be of order $H_{c1}$ to affect the low temperature sense of chirality of the sample.

Circular dichroism and birefringence effects applicable for $p$-wave superconductors were previously calculated by Yip and Sauls \cite{yip}, showing that these effects arise from the order parameter collective mode response of the superconductor. The expression they obtained for frequencies much above the gap frequency is $\theta_K \sim (v_F/c)(\xi/\lambda_L)(\Delta/\epsilon_F)$ln$(\epsilon_F/\Delta)(2\Delta/\hbar \omega)^2$. Here $v_F$ is the Fermi velocity, $\epsilon_F$ is the Fermi energy, $\xi$ is the superconducting coherence length, $\lambda_L$ is the London penetration length, and $\omega = 2\pi c/\lambda$ is the light frequency.  Using known material parameters \cite{mackenzie}, we estimate the effect based on their expression to be $\sim$ 10$^{-12}$ rad.  There could be several reasons for the much larger measured signal and its proportionality to the gap rather than a predicted cubic proportionality.   The multi-band nature of the material, its finite spin-orbit coupling, the fact that $\lambda_L/\xi \approx 2.5$ (Yip and Sauls assumed an extreme type-II limit), or other effects could result in a finite signal even though no effect is expected from the  contribution to the current response from the single-particle excitations, even with particle-hole asymmetry. Alternatively, a larger result proportional to $\Delta$ may be the result of broken TRS that appears below $T_c$, but too close for us to resolve \cite{ohmi}.   While we cannot comment at this point on such possibilities, we attempt a more naive calculation which follows a simple model for a Kerr effect in metallic ferromagnets \cite{erskine}.

Since there are no strong absorption effects in the range of our instrument's wavelength,  the polar Kerr angle is given by \cite{whitegeballe}:  $\theta_K = (4 \pi/\tilde{n}\omega)\sigma_{xy}^{\prime \prime}(\omega)$. Here $\tilde{n} \approx 3$ \cite{katsufuji} is the sample's average index of refraction  and $\sigma_{xy}^{\prime \prime}(\omega)$ is the imaginary part of the off-diagonal component of the conductivity at the light frequency which is given by \cite{erskine}:

\begin{equation}
\sigma_{xy}^{\prime \prime}(\omega) = \frac{\omega_p^2}{4\pi}\frac{|P_0|}{e v_F}\frac{1}{\omega \tau}
\label{imcond}
\end{equation}

\noindent where $\omega_p = 4\pi ne^2/m^*$ is the plasma frequency of the metal, $n$ is the electron density, $m^*$ is the effective mass, $|P_0|$ is the maximum magnitude of the electric dipole moment due to the chiral state, and $\tau$ is the quasiparticle scattering time.  We next assume that the energy associated with $|P_0|$ is proportional to the degree of particle-hole asymmetry  at the Fermi level $\sim\Delta/\epsilon_F$ \cite{ohmi,furusaki}.  Thus, ignoring the multiband nature of \S2RO4 and the specifics of the bands, we expect  $nP_0^2 \sim  \Delta^2/\epsilon_F$.  This implies an approximate  dipole moment  $|P_0| \sim  (\Delta /\epsilon_F) \sqrt{\epsilon_F/n} \sim  (ev_F/\omega_p) (\Delta/\epsilon_F)$, which is proportional to $\Delta$ as is observed experimentally.   Within this simple model, the expected Kerr angle is:

\begin{equation}
\theta_K \sim \frac{\omega_p}{\tilde{n}\omega^2 \tau} \frac{\Delta}{\epsilon_F}
\end{equation}

Using known material parameters \cite{mackenzie}, and  scattering time  $\tau \sim 10^{-11}$ sec (mean free path $\sim$ 1 $\mu$m), extracted from the residual resistivity of similar samples \cite{mackenzie,mao2}, we obtain  $\theta_K \sim$ 100 nanorad. This estimate is very close to the observed saturation signal of $\sim$ 65 nanorad. 

In conclusion, in this paper we showed in an unambiguous way that \S2RO4 breaks time reversal symmetry below the superconducting transition. Coupled with evidence that it is a $p$-wave superconductor \cite{mackenzie}, we believe that we showed that the appropriate order parameter for this system is  of  {\bf d(k)}=$\hat{\bf z}[k_x \pm ik_y]$ variety.  This makes \S2RO4 similar to $^3$He-A \cite{leggett}.

\acknowledgments
Useful discussions with S. Kivelson, I. Mazin, K. Moler, M. Sigrist, and M. Rice are greatly appreciated. We also thank Zhiqiang Mao for his contribution to the crystal growth and to P. SanGiorgio for cryogenic assistance. This work was supported by Center for Probing the Nanoscale, NSF NSEC Grant 0425897 and by the Department of Energy  grant DEFG03-01ER45925.

\end{document}